\documentclass{sample-algotel}
\usepackage[latin1]{inputenc}
\usepackage[francais]{babel}
\usepackage{float}
\usepackage{amsmath}
\usepackage{graphicx}
\usepackage{amssymb}
\usepackage{theorem}
\usepackage{float}
\usepackage{verbatim}

\floatstyle{ruled}
\newfloat{algorithm}{tbp}{loa}
\floatname{algorithm}{Algorithme}

\newtheorem{theo}{Th\'eor\`eme}
\newtheorem{defi}{D\'efinition}
\newenvironment{theorem}[1]{\vspace{-0.35cm}\begin{theo}#1}{\end{theo}\vspace{-0.3cm}}
\newenvironment{definition}[1]{\vspace{-0.35cm}\begin{defi}#1}{\end{defi}\vspace{-0.3cm}}

\author{Shlomi Dolev\addressmark{1}, Swan Dubois\addressmark{2}, Maria Potop-Butucaru\addressmark{2} et S\'ebastien Tixeuil\addressmark{3}}

\title{Communication Optimalement Stabilisante sur Canaux non Fiables et non FIFO}

\address{
\addressmark{1} Universit\'e de Ben-Gurion (Isra\"el), dolev@cs.bgu.ac.il\\
\addressmark{2} UPMC Sorbonne Universit\'es \& INRIA (France), \{swan.dubois,maria.potop-butucaru\}@lip6.fr\\
\addressmark{3} UPMC Sorbonne Universit\'es \& IUF (France), sebastien.tixeuil@lip6.fr
}

\keywords{Canaux non fiables, canaux non FIFO, auto-stabilisation}

\begin{document}
\maketitle

\begin{abstract} 
Un protocole auto-stabilisant a la capacit\'e de converger vers un comportement correct quel que soit son \'etat initial. La grande majorit\'e des travaux en auto-stabilisation supposent une communication par m\'emoire partag\'ee ou bien \`a travers des canaux de communication fiables et FIFO. Dans cet article, nous nous int\'eressons aux syst\`emes auto-stabilisants \`a passage de messages \`a travers des canaux de capacit\'e born\'ee mais non fiables et non FIFO. Nous proposons un protocole de communication (entre voisins) stabilisant et offrant une tol\'erance optimale. Plus pr\'ecis\`ement, ce protocole simule un canal de communication fiable et FIFO garantissant un nombre minimal de pertes, de duplications, de cr\'eations et de r\'e-ordonnancements de messages.
\end{abstract}

\vspace{-0.2cm}
\section{Motivations et D\'efinitions}\label{sec:intro}

L'\emph{auto-stabilisation} \cite{D74j} est une propri\'et\'e des syst\`emes distribu\'es permettant de tol\'erer des fautes transitoires (\emph{i.e.} de dur\'ee finie) de type arbitraire. Plus pr\'ecis\`ement, un syst\`eme est dit auto-stabilisant s'il garantit que toute ex\'ecution issue d'une configuration arbitraire retrouve en un temps fini un comportement conforme \`a la sp\'ecification du syst\`eme et ceci sans aide ext\'erieure (humaine ou autre).

\paragraph{Motivations} \'Etant donn\'e que l'auto-stabilisation est une propri\'et\'e non triviale \`a satisfaire, une large part des travaux traitant de ce domaine se placent dans un mod\`ele de communication tr\`es simple dans lequel tous les processeurs peuvent d\'eterminer de mani\`ere atomique l'\'etat de tous leurs voisins (ce mod\`ele de calcul est connu sous le nom de mod\`ele \`a \'etats). Il est cependant \'evident que ce mod\`ele n'est pas r\'ealiste et qu'un mod\`ele plus classique comme le mod\`ele asynchrone \`a passage de messages est plus proche d'un syst\`eme r\'eel. Dans un tel mod\`ele, les processeurs voisins communiquent par envoi et r\'eception de messages \`a travers le canal de communication qui les s\'epare. Il existe des transformateurs permettant de passer de mani\`ere automatique du premier mod\`ele au second \cite{D00b,DIM93j} ainsi que des algorithmes \'ecrits directement pour le mod\`ele \`a passage de messages \cite{DT06c,BK97j} mais ceux-ci supposent l'existence d'un protocole de communication entre processeurs voisins. Le protocole de communication (entre voisins) le plus connu est le protocole du bit altern\'e (PBA). Il a \'et\'e prouv\'e que ce protocole fournit des propri\'et\'es de stabilisation \cite{AB93j,DIM93j}. En effet, pour toute ex\'ecution du PBA, il existe un suffixe qui satisfait la sp\'ecification (\emph{i.e.} le PBA est pseudo-stabilisant \cite{BGM93j}). Apr\`es les r\'esultats de \cite{GM91j,DIM93j} qui montrent qu'il est impossible de fournir un protocole de communication avec une m\'emoire born\'ee si les canaux sont de capacit\'e non born\'ee, les travaux r\'ecents se sont concentr\'es sur des syst\`emes avec des canaux de capacit\'e born\'ee. Il existe diff\'erents protocoles de communication stabilisants \`a travers des canaux de capacit\'e born\'ee qui diff\'erent par les hypoth\`eses faites sur le syst\`eme (m\'emoire born\'ee, canaux, etc.) mais toutes les solutions connues \cite{BGM93j,HNM99c,V00j} consid\`erent des canaux FIFO.

Un d\'efaut commun \`a tous les protocoles de communication pr\'ec\'edents est qu'ils ne fournissent aucune mesure de l'impact quantitatif des fautes transitoires sur les messages transmis. Partant d'une configuration initiale arbitraire, le contenu initial des canaux est lui aussi arbitraire, ce qui peut conduire le protocole \`a perdre, dupliquer des messages ou bien d\'elivrer de faux messages (qui n'ont pas \'et\'e envoy\'es mais r\'{e}sultent des fautes initiales). Du point de vue de l'application qui utilise le protocole de communication, il est primordial de conna\^itre des bornes sur le nombre de messages pouvant subir de tels al\'{e}as. \`A notre connaissance, seuls \cite{DDNT10j,DT06c} traitent de ce probl\`eme dans une certaine mesure. En effet, ils peuvent \^etre adapt\'es pour obtenir des protocoles de communication instantan\'enement stabilisants. La stabilisation instantan\'ee assure que tout message envoy\'e sera d\'elivr\'e en un temps fini, mais le nombre de messages dupliqu\'es ou de faux messages cr\'ees n'est pas \'etudi\'e. 

\paragraph{Contributions} Notre contribution dans cet article est double. Dans un premier temps, nous d\'efinissons un ensemble de m\'etriques pour mesurer la performance d'un protocole de communication stabilisant et nous donnons des bornes inf\'erieures pour plusieurs d'entre elles. En particulier, nous montrons que tout protocole de communication stabilisant \`a travers des canaux de capacit\'e born\'ee non fiables et non FIFO peut \^etre contraint \`a dupliquer un message, \`a d\'elivrer un faux message ou \`a r\'e-ordonner un message. Dans un second temps, nous proposons un protocole optimal par rapport aux bornes inf\'erieures \'evoqu\'ees pr\'ec\`edement.

\paragraph{Sp\'ecification} Nous consid\'erons ici un syst\`eme distribu\'e \`a passage de messages r\'eduit \`a deux processeurs : $p_i$ qui sera consid\'er\'e comme \'emetteur de messages et $p_j$ qui sera consid\'er\'e comme r\'ecepteur de messages. Le canal de communication s\'eparant $p_i$ et $p_j$ est constitu\'e de deux canaux virtuels de directions oppos\'ees. Le premier, $(i,j)$, permet \`a $p_i$ d'envoyer des messages \`a $p_j$ tandis que le second, $(j,i)$, permet \`a $p_j$ d'envoyer des acquittements \`a $p_i$. Chacun de ces canaux virtuels est asynchrone (le temps de livraison de tout message est fini mais non born\'e), a une capacit\'e born\'ee de $c$ messages (tout envoi de messages lorsque cette borne est atteinte conduit \`a la perte d'un message arbitraire), non fiable (tout message peut \^etre perdu \`a un moment arbitraire) mais \'equitable (tout message envoy\'e infiniment souvent est re\c cu infiniment souvent) et non-FIFO (l'ordre d'arriv\'ee des messages est ind\'ependant de l'ordre d'envoi). Il faut noter que, en raison du contexte auto-stabilisant, chaque canal virtuel contient initialement jusqu'\`a $c$ messages de contenu arbitraire.

La sp\'ecification que nous pr\'esentons \`a pr\'esent est inspir\'ee de celle de \cite{L96b} mais elle est adapt\'ee au contexte auto-stabilisant. Supposons que nous avons une application distribu\'ee qui souhaite envoyer des messages de $p_i$ \`a $p_j$. Notre objectif est de fournir un protocole de communication \`a cette application qui remplit cette t\^ache de mani\`ere transparente malgr\'e les caract\'eristiques du canal de communication. Cette application \emph{envoit} un message lorsqu'elle demande au protocole de communication de faire parvenir un message depuis $p_i$ vers $p_j$. Un message est \emph{d\'elivr\'e} \`a $p_j$ lorsque le protocole de communication fournit ce message \`a l'application s'ex\'ecutant sur $p_j$. Un message \emph{fant\^ome} est un message d\'elivr\'e \`a $p_j$ alors qu'il n'a pas \'et\'e envoy\'e par $p_i$. Un message \emph{dupliqu\'e} est un message d\'elivr\'e plusieurs fois \`a $p_j$ alors qu'il n'a \'et\'e envoy\'e qu'une fois par $p_i$. Un message \emph{perdu} est un message envoy\'e par $p_i$ mais jamais d\'elivr\'e \`a $p_j$. Un message $m$ est \emph{r\'e-ordonn\'e} lorsqu'il d\'elivr\'e \`a $p_j$ avant un message $m'$ alors que $m$ a \'et\'e envoy\'e apr\`es $m'$ par $p_i$. Le but d'un protocole de communication stabilisant est alors de fournir des propri\'et\'es sur le nombre de messages perdus, dupliqu\'es, fant\^omes et r\'e-ordonn\'es. Nous sp\'ecifions notre probl\`eme comme suit :

\begin{definition}
Un protocole de communication est $(\alpha,\beta,\gamma,\delta)$-stabilisant sur des canaux $c$-born\'es s'il remplit les conditions suivantes pour toute ex\'ecution issue d'une configuration arbitraire:\\
- Dans le pire cas, seuls les $\alpha$ premiers messages envoy\'es par $p_i$ peuvent \^etre perdus.\\
- Dans le pire cas, seuls les $\beta$ premiers messages d\'elivr\'es \`a $p_j$ peuvent \^etre des messages dupliqu\'es.\\
- Dans le pire cas, seuls les $\gamma$ premiers messages d\'elivr\'es \`a $p_j$ peuvent \^etre des messages fant\^omes.\\
- Dans le pire cas, seuls les $\delta$ premiers messages d\'elivr\'es \`a $p_j$ peuvent \^etre des messages r\'e-ordonn\'es.
\end{definition}

\vspace{-0.2cm}
\section{Bornes inf\'erieures}\label{sec:bornes}

\'Etant donn\'e que tout protocole de communication doit avoir dans son code une instruction pour d\'elivrer les messages \`a l'application et que, dans un contexte auto-stabilisant, le compteur ordinal peut \^etre arbitrairement corrompu dans la configuration initiale, il est possible que la premi\`ere instruction execut\'ee par le processeur r\'ecepteur soit la livraison d'un message qui n'a jamais \'et\'e envoy\'e, \emph{i.e.} un message fant\^ome. Si, de plus, ce message fant\^ome est identique \`a un autre message envoy\'e par $p_i$ dans l'ex\'ecution consid\'er\'ee, ce message peut devenir un message dupliqu\'e ou r\'e-ordonn\'e. Nous obtenons les r\'esultats suivants.

\begin{theorem}
Il n'existe pas de protocole de communication $(\alpha,\beta,\gamma,\delta)$-stabilisant sur des canaux $c$-born\'es avec $\beta=0$, $\gamma=0$ ou $\delta=0$.
\end{theorem}

\vspace{-0.2cm}
\section{Protocole de communication $(0,1,1,1)$-stabilisant}\label{sec:algo}

Nous sommes maintenant en mesure de pr\'esenter notre protocole de communication. Celui-ci est compos\'e de deux fonctions : \textbf{Send($m$)} qui est ex\'ecut\'ee par $p_i$ \`a chaque fois qu'il souhaite envoyer un message $m$ (\textbf{Send} est bloquant, \emph{i.e.} $p_i$ doit attendre la fin de son ex\'ecution avant de commencer \`a envoyer le message suivant) et \textbf{Receive()} qui est ex\'ecut\'ee par $p_j$ en continu. 

\paragraph{Id\'ee g\'en\'erale} L'id\'ee de base de notre protocole est de modifier le PBA de mani\`ere \`a am\'eliorer ses propri\'et\'es de tol\'erance. Si le processeur $p_i$ souhaite envoyer un message $m$, il envoie celui-ci de mani\`ere p\'eriodique et $p_j$ acquitte chaque copie de $m$ qu'il re\c coit. Le processeur $p_j$ n'est autoris\'e \`a d\'elivrer le message $m$ que lorsqu'il en a re\c cu $c+1$ copies (pour assurer qu'au moins l'une d'entre elles a bien \'et\'e envoy\'ee par $p_i$). De plus, $m$ n'est d\'elivr\'e que si la valeur du bit altern\'e qui lui est associ\'ee est diff\'erente de celle du dernier message d\'elivr\'e par $p_j$ (de mani\`ere \`a assurer que le message ne soit pas dupliqu\'e car $p_i$ continue d'envoyer $m$ tant qu'il n'a pas re\c cu suffisament d'acquittements). Afin d'assurer que $p_j$ a re\c cu au moins $c+1$ copies du message, $p_i$ attend d'avoir re\c cu $3c+2$ acquittements avant d'arr\^eter d'envoyer $m$ (en effet, au plus $c+1$ acquittements sont d\^us \`a la configuration initiale tandis que au plus $c$ sont d\^us \`a la pr\'esence initiale de messages erronn\'es dans le canal $(i,j)$). \`A ce stade, notre protocole ne garantit pas encore l'absence de pertes de messages \`a cause de l'utilisation du bit altern\'e (en effet, si le bit altern\'e du message et du r\'ecepteur ne sont pas initialement synchronis\'es, le premier message envoy\'e par $p_i$ peut \^etre perdu). Pour \'eviter cela, $p_i$ alterne entre l'envoi de messages de synchronisation et de $m$. Plus pr\'ecis\`ement, pour envoyer un message $m$, $p_i$ commence par envoyer un message de synchronisation (not\'e $<SYNCHRO>$) jusqu'\`a recevoir $3c+2$ acquittements avant d'envoyer le message $m$ lui-m\^eme jusqu'\`a en recevoir $3c+2$ acquittements. Il s'ensuit que seul le message de synchronisation est perdu dans le pire des cas.

\paragraph{Pr\'esentation d\'etaill\'ee} Notre protocole est pr\'esent\'e en Figure 1. La proc\'edure \textbf{Send} se contente d'envoyer un message de synchronisation puis le message re\c cu en param\`etre (\`a l'aide de la fonction auxilliaire \textbf{SendMessage}) apr\`es avoir altern\'e la valeur du bit associ\'e. Finalement, elle d\'elivre un acquittement \`a l'application \`a l'aide de la fonction \textbf{DeliverAck}. La fonction auxilliaire \textbf{SendMessage} envoit p\'eriodiquement le message \`a l'aide de la fonction \textbf{SendPacket} (qui permet d'envoyer un paquet sur le canal $(i,j)$) et compte le nombre d'acquittement re\c cus en faisant appel \`a la fonction \textbf{ReceivePacket} (qui permet de r\'ecup\'erer un message dans le canal $(j,i)$). Celle-ci s'arr\^ete lorsqu'elle a compt\'e $3c+2$ acquittements.

\begin{figure*}[htb!]
\scriptsize
\centering
\begin{tabular}{|p{2.75in}||p{2.75in}|}
\hline
\begin{minipage}[t]{2.75in}
\centering
{\it\bf Send}
\begin{tabbing}
X: \= d \= d \= d \= d \= d \= d \= d \= d \= \kill

\textbf{entr\a'ee:} \>\>\>\>\>
$m$: message \a`a envoyer\\

\textbf{variable:} \>\>\>\>\>
$ab$: bool\a'een donnant la valeur du bit altern\a'e actuelle\\

01: \>\>$ab := \neg ab$ \\
02: \>\>\textbf{SendMessage} $(<SYNCHRO>,ab)$ \\
03: \>\>$ab := \neg ab$ \\
04: \>\>\textbf{SendMessage} $(m,ab)$ \\
05: \>\>\textbf{DeliverAck} ($m$) \\
\end{tabbing}
\vspace{-0.25cm}
\centering
{\it\bf SendMessage}
\begin{tabbing}
X: \= d \= d \= d \= d \= d \= d \= d \= d \= \kill

\textbf{entr\a'ee:}\>\>\>\>\> 
$m'$: message \a`a envoyer\\
\>\>\>\>\> $ab$: bool\a'een donnant la valeur du bit altern\a'e associ\a'e \a`a $m$\\

\textbf{variable:}\>\>\>\>\>
$ack$: entier donnant le nombre d'acquittements re\c cu pour\\ 
\>\>\>\>\>\>\>\>la valeur actuelle de $ab$\\

01: \>\>$ack := 0$\\
02: \>\>\emph{while} $ack < 3c+2$ \\
03: \>\> \>\textbf{SendPacket} $(m',ab)$ \\
04: \>\> \>\emph{if} \textbf{ReceivePacket} $(ack,(m',ab))$ \\
05: \>\> \> \>$ack := ack+1$; 

\end{tabbing}
\end{minipage}
&
\begin{minipage}[t]{2.75in}
\centering
{\it\bf Receive}
\begin{tabbing}
X: \= d \= d \= d \= d \= d \= d \= d \= d \= \kill 

\textbf{variables:}\\
$last\_delivered$: bool\a'een donnant la valeur du bit altern\a'e du\\
\>\>\>\>\>\>\>\> dernier message d\a'elivr\a'e\\
$Q$: file de taille $c+1$ de $3$-tuples $(m,ab,count)$, o\a`u $m$ est un message,\\
\> $ab$ est une valeur du bit altern\a'e, et $count$ est un entier donnant le\\
\>  nombre de paquets $(m,ab)$ re\c cus pour $m$ et $ab$ depuis le dernier\\
\>  \textbf{DeliverMessage} ou \textbf{DropMessage}. L'op\a'erateur $[]$ renvoit un\\
\> pointeur sur le $count$ associ\a'e \a`a son param\a`etre et place ce 3-tuple\\
\>  en t\a^ete de liste\\

01: \>\>\emph{upon} \textbf{ReceivePacket} $(m,ab)$ \\
02: \>\> \>$Q[m,ab] := min(Q[m,ab]+1,c+1)$\\
03: \>\> \>\emph{if} $Q[m,ab] \geq c+1$ \emph{then} \\
04: \>\> \> \>\emph{if} $last\_delivered \neq ab$ \emph{then}\\ 
05: \>\> \> \> \>\emph{if} $m\neq<SYNCHRO>$ \emph{then}\\
06:\> \> \> \> \> \>\textbf{DeliverMessage} ($m$)\\
07: \>\> \> \> \>\emph{else}\\
08:\> \> \> \> \> \>\textbf{DropMessage} ($m$)\\
09:\> \> \> \> \>$last\_delivered := ab$\\
10: \>\> \> \>$Q := \bot$ \\
11: \>\> \>\textbf{SendPacket} $(ack,(m,ab))$
\end{tabbing}
\end{minipage}\\[1ex]
\hline
\end{tabular}
\normalsize
\caption{$\mathcal{SDL}$, un protocole de communication $(0,1,1,1)$-stabilisant.} 
\label{algo:SDL}
\end{figure*}

\`A chaque r\'eception de message (r\'ealis\'ee gr\^ace \`a la fonction \textbf{ReceivePacket}), la proc\'edure \textbf{Receive} incr\'emente le compteur associ\'e au message qu'elle vient de recevoir. Dans le cas o\`u le message a \'et\'e re\c cu $c+1$ fois, la file servant \`a stocker les messages re\c cus est vid\'ee. Si, de plus, la valeur du bit altern\'e est diff\'erente de celle du dernier message re\c cu au moins $c+1$ fois, alors le message est soit d\'elivr\'e \`a l'application \`a l'aide de \textbf{DeliverMessage} (s'il s'agit d'un message normal) ou bien d\'etruit \`a l'aide de la fonction \textbf{DropMessage} (s'il s'agit d'un message de synchronisation qui est donc sans int\'er\^et pour l'application). Dans les deux cas, le bit du r\'ecepteur est altern\'e. Tout message re\c cu est acquitt\'e (\`a l'aide de la fonction \textbf{SendPacket}) avant de traiter le suivant.

\paragraph{Propri\'et\'es} Nous avons vu pr\'ec\'edement que $p_i$ attend de recevoir $3c+2$ acquittements de chaque message pour arr\^eter de l'envoyer, ce qui garantit que $p_j$ a re\c cu au moins $2c+2$ copies de ce message (dont au moins $c+1$ r\'eellement envoy\'ees par $p_i$) et donc que ce message a bien \'et\'e d\'elivr\'e \`a $p_j$ si $ab\neq last\_delivered$. Si ce n'est pas le cas, l'usage du message de synchronisation nous garantit que notre protocole ne perd aucun message envoy\'e par $p_i$. L'usage du bit altern\'e nous garantit l'absence de duplication apr\`es la premi\`ere r\'eception (si le premier message re\c cu est un message fant\^ome, celui-ci peut \^etre la copie d'un message valide ult\'erieur, ce qui cause au pire une duplication). Le fait d'attendre de recevoir $c+1$ copies de chaque message avant de le d\'elivrer garantit que seul le premier message d\'elivr\'e peut \^etre un message fant\^ome. Enfin, le fait qu'un message $m$ soit d\'elivr\'e \`a $p_j$ entre le d\'ebut et la fin de l'ex\'ecution de \textbf{Send(m)} par $p_i$ et que les appels \`a cette fonction soient bloquants pour $p_i$ implique que seul le premier message peut \^etre r\'e-ordonn\'e (si le premier message re\c cu est un message fant\^ome, celui-ci peut \^etre la copie d'un message valide ult\'erieur, ce qui cause au pire un r\'e-ordonnancement). En conclusion, nous avons le r\'esultat suivant\footnote{Une preuve compl\`ete de ce r\'esultat peut \^etre trouv\'ee dans \cite{DDPT10r}.} : 

\begin{theorem}
$\mathcal{SDL}$ est un protocole de communication $(0,1,1,1)$-stabilisant \`a travers des canaux de communication de capacit\'e born\'ee mais non fiables et non FIFO.
\end{theorem}

\vspace{-0.2cm}
\section{Conclusion}\label{sec:conclusion}

Dans cet article, nous avons introduit des mesures de l'effet de fautes transitoires sur les performances des protocoles de communication entre voisins dans un syst\`eme \`a passage de messages. Nous avons ensuite fourni un protocole optimal par rapport \`a ces mesures dans le cas o\`u les canaux de communications ont une capacit\'e born\'ee, sont non fiables et non FIFO. Toutefois, notre protocole induit un surco\^ut de communication~; la question de savoir s'il est possible de conserver cette tol\'erance optimale aux fautes transitoires en baissant ce surco\^ut de communication de mani\`ere significative est toujours ouverte. 

\vspace{-0.2cm}
\bibliographystyle{alpha}
\small{
\bibliography{Biblio}
}

\end{document}